\RequirePackage{fix-cm}
\documentclass[smallextended]{svjour3}       
\smartqed  
\usepackage{graphicx}

\usepackage{tikz-feynman}
\usepackage{graphicx,psfrag,subfigure}
\usepackage[margin=0.4cm]{caption}

\def\beq{\begin{eqnarray}}
	\def\eeq{\end{eqnarray}}
\def\bea{\begin{eqnarray*}}
	\def\eea{\end{eqnarray*}}

\def\half{{1\over 2}}

\begin{document}
\title{Supersymmetry : A decade after Higgs discovery
}


\author{V. Suryanarayana Mummidi   \and     Priyanka Lamba \and Sudhir K. Vempati}


\institute{V. Suryanarayana Mummidi \at
              Department of Physics, National Institute of Technology, Tiruchirappalli 620015 \\
              \email{venkata@nitt.edu}           
           \and
           Priyanka Lamba \at
              IFT, Department of Physics, University of Warsaw, Poland \\ 
              \email{priyanka.lamba@fuw.edu.pl}
            \and 
            Sudhir K Vempati \at 
            Centre for High Energy Physics, Indian Institute of Science, Bangalore 560 012 \\ 
            \email{vempati@iisc.ac.in}
}

\date{Received: date / Accepted: date}

\maketitle

\begin{abstract}

Supersymmetric extensions of the Standard Model have been in vogue for over half a century. They have many interesting theoretical properties like calculability, absence of quadratic divergences 
, and phenomenologically impactful features like gauge coupling unification, dark matter candidates,
signatures at present and future colliders, etc. A defining feature of these models is the 
calculability of Higgs mass in terms of a few parameters.  The discovery of a Higgs particle 
with a mass of around 125 GeV  thus has  significant
implications. The null results for the searches of superpartners at LHC has also put further constraints. Taken together with direct detection limits on WIMP (Weakly Interacting Massive Particle)
dark matter, it appears that TeV scale supersymmetry is not realised in Nature and the theoretical expectations have reached a turning point. The present onslaught from the experiments suggests
that supersymmetric models  need a more complex particle structure, lagrangian and breaking patterns to be a natural solution to the hierarchy problem. 
We review existing  models and discuss their feasibility in the current and future experimental programs.

\end{abstract}



\section{Introduction}

Supersymmetry was once considered as the pinnacle  of physics beyond Standard Model (BSM). Since it has been discovered in the early
$'70$s,  it's several virtues  made it a favourite BSM model  amongst theorists, phenomenologists and experimentalists all alike. 
For recent reviews of supersymmetry, please see, \cite{Bertolini,Luty:2005sn,Bilal:2001nv,Weinberg:2000cr,Martin:1997ns} and references there in. 
Amongst the several virtues listed for supersymmetry, in particular for the Minimal Supersymmetric Standard Model (MSSM) are: (a) a calculable and renormalisable theory protected by non-renormalisation theorems (b) a symmetry to protect the Higgs mass
(c) gauge coupling unification (d) a natural candidate for dark matter if R-parity is conserved and (e) a dynamical understanding of electroweak symmetry breaking  and so on. 

While these virtues made it popular, however the lack of any signal either in direct searches at colliders like LEP, Tevatron, even LHC (during the first two runs) or in indirect searches like flavor and CP violation in charged leptonic\footnote{There are some hints of new physics in the muon $g-2$ experiment, which we will discuss it later in the text.}, B or K meson systems,  EDMs (electric Dipole Moments) and in direct detection experiments for Weakly Interacting Massive Particle (WIMP) dark matter experiments, make weak scale supersymmetry $\sim 100$ GeV - TeV  severely disfavored scenario. 
In the present review, we revisit the current status of low energy supersymmetry including constraints from Higgs mass, flavour and CP, colliders and dark matter. Along the way, we discuss the recent supersymmetric explanations for the muon $g-2$ anomaly. Taking the viewpoint that supersymmetry might still exist in some form, close to the electroweak scale, we look at the possible spectra that come out after considering all the constraints. While all the supersymmetric particles might not completely exist in the TeV corridor, it could be that only a part of them could exist in that corridor. For example, weakly interacting fermionic superpartners could still lie in that energy regime, while the rest could be distributed upwards in mass. We then discuss about the present status of some popular supersymmetric breaking models importantly gauge mediation, minimal supergravity/CMSSM and so on. 

The organization of the review is as follows: In the next section, we review all the phenomenological constraints on supersymmetric models in various subsections. In section 3, we look at possible spectra of supersymmetric particles as an outcome of all the present constraints and survey some models which can lead to them. In section 4, we look at present situation of the various models of supersymmmetry breaking and novel scenarios leading to naturally heavy supersymmetric spectra. We finally close with an outlook and summary in section 5. 

\section{Phenomenological Constraints on MSSM} 
In the present section we review the various phenomenological constraints on supersymmetric theories mainly focusing on MSSM. The constraints can be divided as direct, mostly coming from collider (LHC) limits, indirect (flavour and CP) and astrophysical (dark matter relic density and direct detection ). However, one of the main features of the MSSM is the calculability of the Higgs mass in terms of the other fundamental parameters of the model. The discovery of the Higgs particle with its mass reasonably precisely measured thus puts strong constraints on the supersymmetric parameter space. The measurement of the Higgs mass strongly disfavors many supersymmetric models. For this reason we start the present section with a review of the Higgs discovery, and it's implications for the supersymmetric models. 

\subsection{Implications of Higgs Boson discovery. }
On July 4, 2012, about a decade ago, both the LHC experiments CMS and ATLAS announced the discovery of a Higgs-like particle with a mass of around 126 GeV. Over the decade, Higgs has been seen with more statistics, final states, and production mechanisms, leading to a precise value of its mass and many of its couplings (see, for example, \cite{ATLAS:2022vkf,CM022dwd}). At present, the Higgs mass is measured to be \cite{ParticleDataGroup:2022pth} 
\begin{equation}
  \label{higgsexpt}
 m_h = 125.25 \pm 0.17\, \rm{GeV}.
\end{equation}

The MSSM as well known as two Higgs doublet model with only one quartic coupling determined by the gauge couplings. This leads to a simple prediction for the 
mass of the CP-even neutral Higgs bosons at the tree level (see for example, discussion in \cite{Martin:1997ns,Drees:2004jm}). 
\begin{eqnarray}
m^2_{h, H} &=& \half
\Bigl (
m^2_{A} + m_Z^2 \sqrt{(m_{A}^2 - m_Z^2)^2 + 4 m_Z^2 m_{A}^2 \sin^2 (2\beta)} 
\Bigr ).
\label{cpevenhiggs}
\end{eqnarray}
here we set the neutral Higgs vacuum expectation values in CP conserving MSSM as
$v_u = \langle H_u^0\rangle$, $ v_d = \langle H_d^0\rangle$ with the ratio $\tan\beta = \frac{v_u}{v_d}$, $H_u^0$ ($H_d^0$) is the Higgs which gives masses for the up type (down type) fermions. 
The heavy (light) CP-even neutral scalars are written as $H$ ($h$), while $ A$ denotes the CP-odd neutral scalar. It is easy to see that the mass of the lightest Higgs boson is bounded from above at the tree level as 
\begin{equation}
\label{higgsmassrelations}
    m_h \leq~ M_Z |\cos 2 \beta|. 
\end{equation}
This is the so-called tree-level Higgs `catastrophe', where the computed lightest neutral CP even Higgs has a maximum mass of the $Z$ boson.  This catastrophe is solved by loop corrections as has been shown by various authors in the early nineties (See for example, \cite{Haber:1990aw}). At 1-loop, corrections are proportional to the top Yukawa coupling and thus can be large. We briefly summarise the result below. 

Using the well-known effective potential method, the 1-loop corrections to the CP-even part of the Higgs mass matrix at zero momentum in $\overline{DR}$ scheme is :
\begin{equation}
M^2_{Re} = M^2_{Re}(0) + \delta M^2_{Re},
\end{equation}
where $M^2_{Re}(0)$ represents the tree level mass matrix given by eq.(\ref{cpevenhiggs}) and $\delta M^2_{Re}$ represents its one-loop corrections. The dominant one-loop corrections come from the top quark and stop squark loops which can be written in the following form \cite{Drees:2004jm}:
\begin{equation}
\delta M^2_{Re} = \left( \begin{array}{cc}
\Delta_{11} & \Delta_{12} \\
\Delta_{12} & \Delta_{22} 
\end{array} \right),
\end{equation}
where 
\begin{eqnarray}
\Delta_{11} &=& {3 G_F\, m_t^4 \over 2\sqrt{2} \pi^2 \sin^2 \beta} 
\left[ { \mu ( A_t + \mu \cot \beta) \over m_{\tilde{t}_1}^2 - m_{\tilde{t}_2}^2} \right]^2 
\left( 2 - {m_{\tilde{t}_1}^2 + m_{\tilde{t}_2}^2 \over m_{\tilde{t}_1}^2 - m_{\tilde{t}_2}^2 }
 \ln {m_{\tilde{t}_1}^2 \over 
m_{\tilde{t}_2}^2} \right) \nonumber \\
\Delta_{12} &=& {3 G_F\, m_t^4 \over 2\sqrt{2} \pi^2 \sin^2 \beta} 
\left[ { \mu ( A_t + \mu \cot \beta) \over m_{\tilde{t}_1}^2 - m_{\tilde{t}_2}^2} \right]
 \ln {m_{\tilde{t}_1}^2 \over m_{\tilde{t}_2}^2} + {A_t \over \mu} \Delta_{11} \nonumber \\
\Delta_{22} &=& {3 G_F\, m_t^4 \over \sqrt{2} \pi^2 \sin^2 \beta} 
\left[  \ln {m_{\tilde{t}_1}^2 m_{\tilde{t}_2}^2  \over m_t^2} +
{A_t (A_t +  \mu \cot \beta) \over  m_{\tilde{t}_1}^2 -  m_{\tilde{t}_2}^2} 
 \ln {m_{\tilde{t}_1}^2 \over m_{\tilde{t}_2}^2} \right]
+ \left({A_t \over \mu}\right)^2 \Delta_{11}. 
\end{eqnarray}
In the above $G_F$ represents Fermi decay constant, $m_t$ the top mass, $m_{\tilde{t}_1}^2 $ and $m_{\tilde{t}_2}^2$ are the eigenvalues of the stop mass matrix, $A_t$ is the trilinear scalar coupling (corresponding to the top Yukawa coupling) in the stop mass matrix and $\mu$ is supersymmetric Higgs mass parameter. Taking into account these corrections, the condition (\ref{higgsmassrelations}) takes the form:
\begin{equation}
\label{higgs1loop}
m_h^2~ < ~ m_Z^2\, \cos^2( 2 \beta) + \Delta_{11} \,\cos^2 \,\beta + \Delta_{12}\, \sin 2 \beta + \Delta_{22}\,
\sin^2 \beta,
\end{equation}
thus circumventing the tree-level catastrophe. 

In terms of effective field theory computation, the one-loop contribution can be estimated as a
correction to the quartic term of the Higgs potential, leading to a formula for the one-loop correction as follows. It should be noted that this formula works very well reproducing the features of the full computation. 

\begin{equation}
m_h^2\le m_Z^2\cos^2(2\beta) + \frac{3g^2m_t^4}{8\pi^2m_W^2}\left[\ln{\frac{M_S^2}{m_t^2}}+\frac{X_t^2}{M_S^2}\left(1-\frac{X_t^2}{12M_S^2}\right)\right],
\end{equation}
where $M_S$ is an average stop mass, $X_t \equiv A_t + \mu \cot \beta$ is the trilinear coupling (top-squark mixing parameter), and $m_Z^2\cos^2(2\beta)$ is tree level contribution in MSSM. 
Over the years, the calculations for the Higgs mass have been improved beyond the one loop order with leading two-loop corrections, effective field theory including resummations etc. For recent reviews 
please see \cite{Draper:2016pys}. In this review, we quote the results from the recent summary of
Ref. \cite{Slavich:2020zjv}.  The computations are included in the recent version of FeynHiggs \cite{Bahl:2018qog}. 
\begin{figure}[h!]
\centering    
\subfigure{\includegraphics[width=8cm,height=7cm]{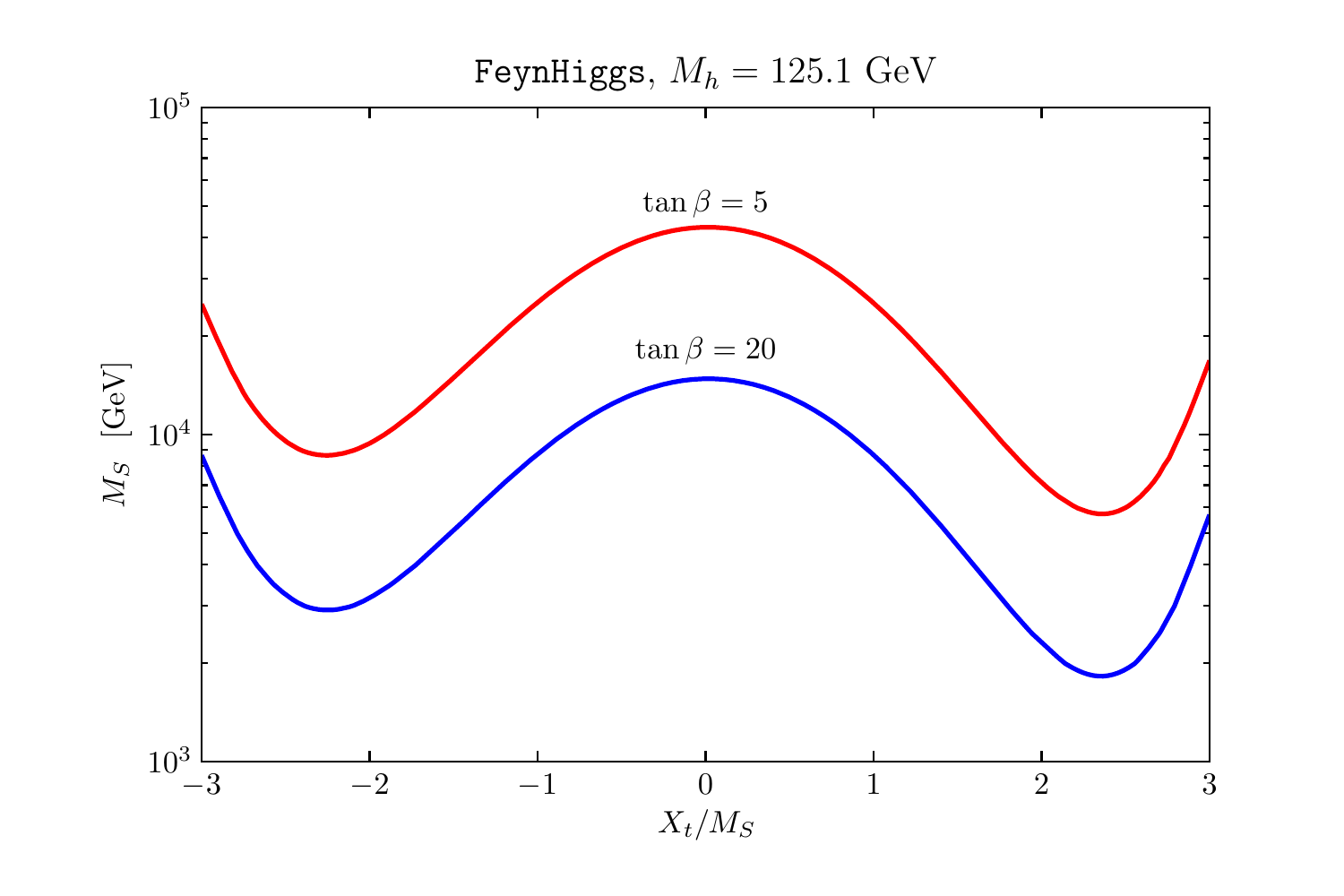}}
\subfigure{\includegraphics[width=8cm,height=7cm]{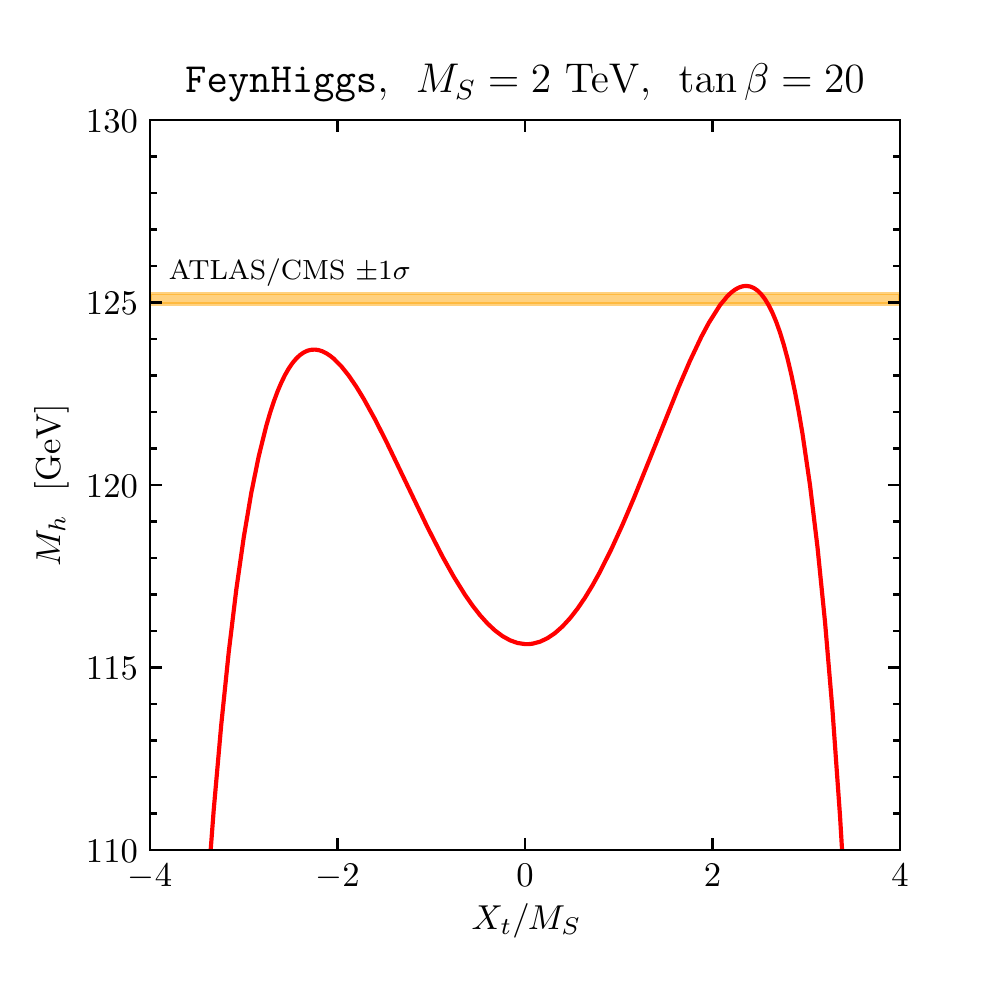}}
\caption{The left panel shows $M_S$ vs $X_t/M_S$ plane for two values of $\tan\beta$ with correct Higgs mass. The right panel  shows the Higgs mass vs $X_t$ plane for $M_S \sim 2 $ TeV and $\tan\beta = 20.$}
\label{mhvsxt}
\end{figure}
From the Fig \ref{mhvsxt}, we see that the measured value of Higgs as in eq.(\ref{higgsexpt})  would require either a large mass of stop of around 10 TeV \cite{Kitano:2006gv,Bagnaschi:2018igf} or large trilinear coupling for lighter stop mass (of less than 2-3 TeV). 

\begin{figure}[h!]
\centering    
\subfigure{\includegraphics[width=0.46\textwidth]{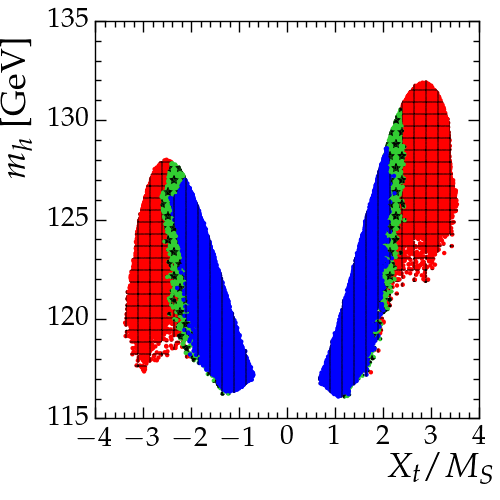}}
\subfigure{\includegraphics[width=0.46\textwidth]{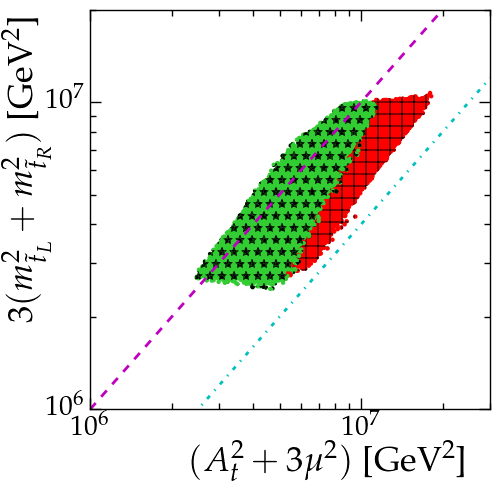}}
\caption{ The red, green and blue regions correspond to deeper charge and colour-breaking (CCB) minima, metastable Higgs vacua and to stable Higgs vacua respectively. The left panel shows the relevant regions in Higgs mass vs $X_t/M_S$ plane and the right panel shows in $A_t^2+\mu^2$ vs $3(m_{\tilde{t}_L}^2 + m_{\tilde{t}_R}^2$) plane.}
\label{stopccb}
\end{figure}

The stop mixing can be subjected to charge and colour breaking (CCB) minima which can put strong constraints on the stop mixing parameters and their masses \cite{Chowdhury:2013dka}. The regions are shown in Fig \ref{stopccb}, where CCB minima could lead to unstable (deeper global) or metastable  minima compared to the preferred Higgs vacua. The metastable regions are characterised by lifetimes which are equal to or longer than the age of the universe. 

\subsection{Limits from LHC}
\begin{figure}[h!]
\centering   
\subfigure{\raisebox{2.5cm}{\includegraphics[scale=0.8]{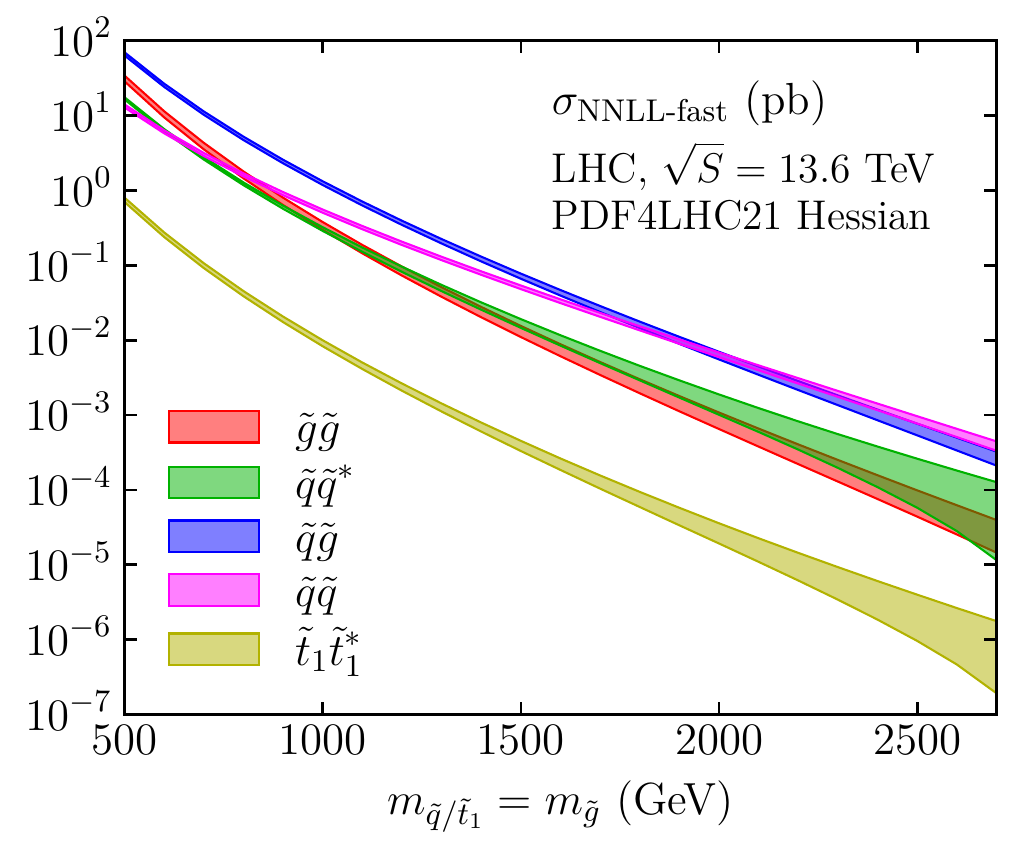}}}
\subfigure{\includegraphics[scale=0.8]{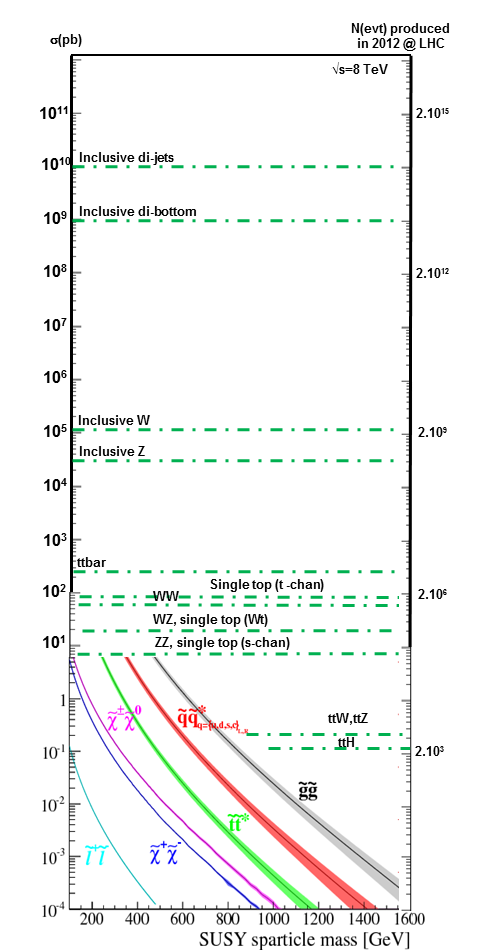}}
\caption{The NLO production crosssections of SUSY particles at the LHC for $\sqrt{s} = 13.6$ TeV in the left panel \cite{Borschensky:2014cia}. Typical background rates are presented in the right panel.}
\label{production}
\end{figure}

At the LHC, the dominant processes are strong processes, which lead to the production of coloured  
supersymmetric particles, such as gluinos and squarks. The main production channels are through
$qq$, $qg $, and $gg$ initial states. The production cross-sections
are large  about 1pb for the first two generations of squarks and gluinos of masses around a TeV.  
The cross-sections however fall off rapidly 
with increasing masses as shown in Fig. \ref{production}.  As can be seen in the figure, the 
production cross-sections for stops are  about an order of magnitude smaller for 100 GeV
stops, but fall even more rapidly reaching $\sim$ 10 fb for 1 TeV stops. The backgrounds are very large, typically by
several orders of magnitude as shown in the right panel of Fig. \ref{production}.  In spite of these difficulties, the LHC experiments, 
ATLAS and CMS looking for supersymmetry and have already put strong constraints on the masses of the 
superpartners.

\begin{figure}[h!]
\centering    
\subfigure{\includegraphics[width=0.46\textwidth]{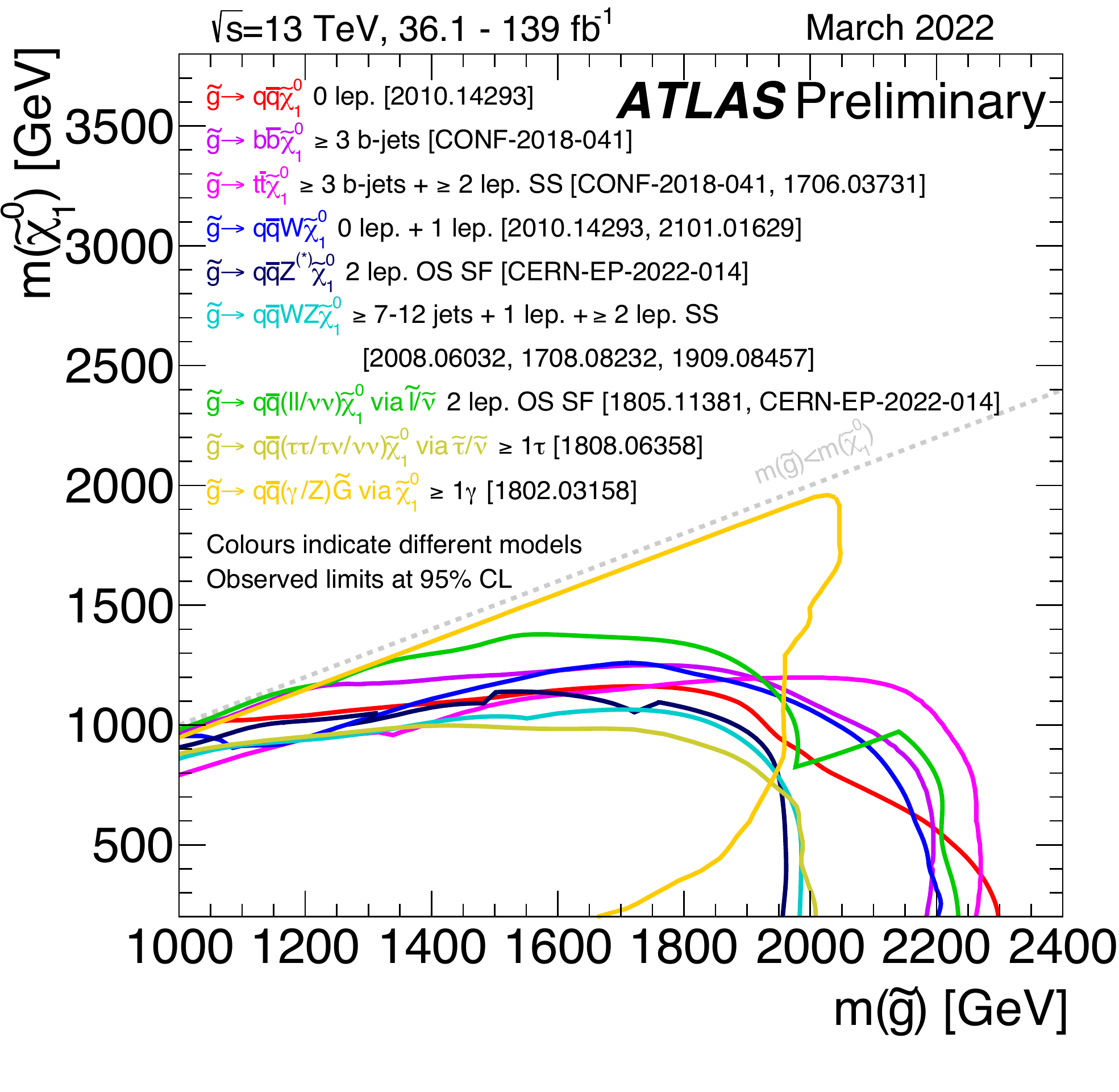}}
\subfigure{\includegraphics[width=0.46\textwidth]{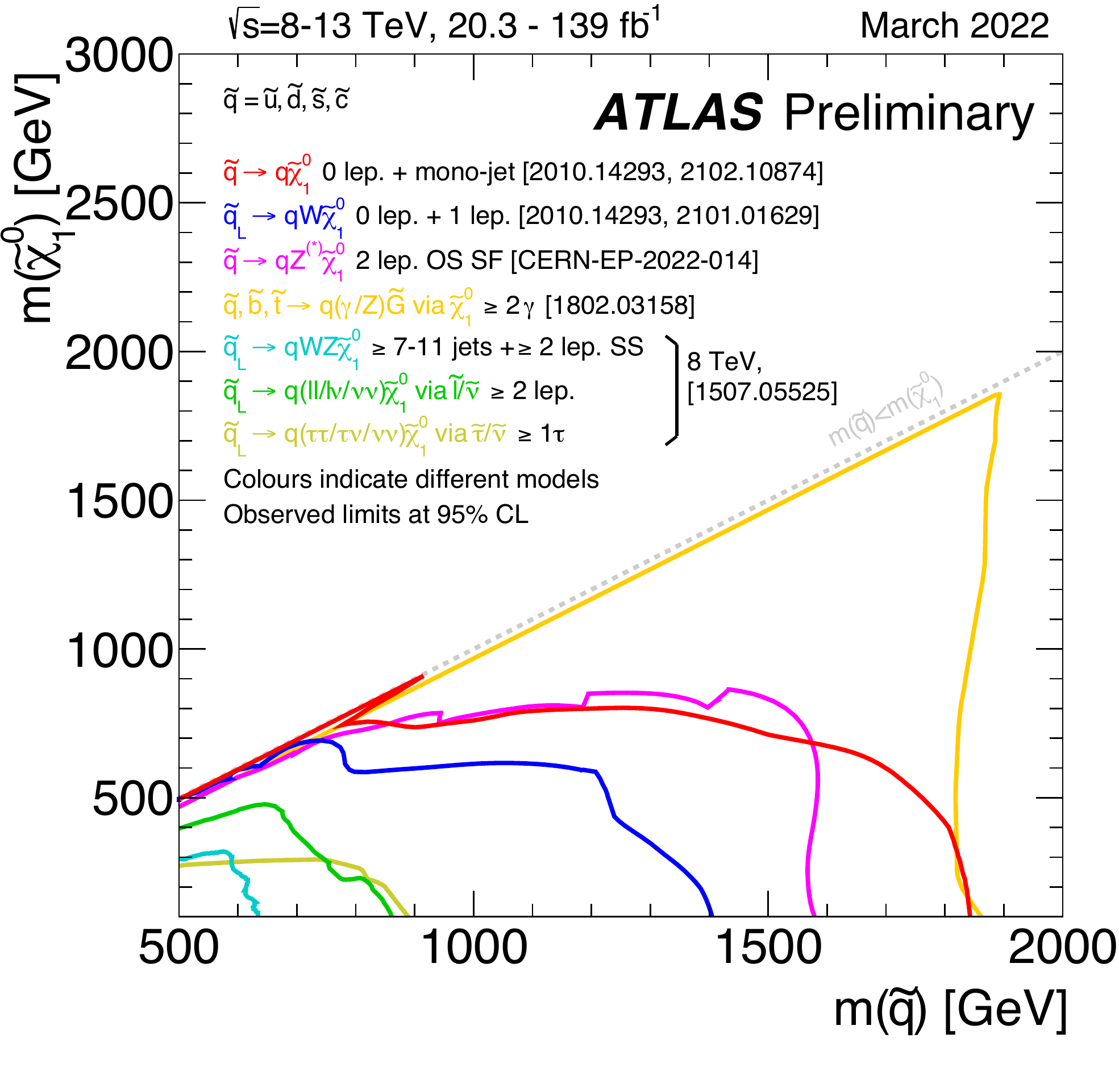}}
\caption{ATLAS collaboration's latest limits on gluino (left panel) and squark (right panel) masses from LHC. The limits are derived within the simplified class of models.}
\label{squarks and gluinos}
\end{figure}

Below we present the limits from the ATLAS Collaboration \cite{ATLASSUSY}.
As expected the strongest constraints are on the coloured supersymmetric partners such as gluinos and
first two-generation squarks. Gluinos are ruled out between 0.8-2.3 TeV depending on the lightest neutralino
mass. This can be seen from the left panel of  Figure \ref{squarks and gluinos}. The limit
on the gluino mass is one of the strongest on the strongly interacting particles of the MSSM. This has implications on most models of supersymmetry with gaugino mass unification, where the low energy masses of gluino, bino and wino are all correlated. The first two generation squarks also have similar limits from ATLAS and CMS. They limits can be seen on the right panel of \ref{squarks and gluinos}. 

\begin{figure}[h!]
\centering    
\subfigure{\includegraphics[width=0.55 \textwidth]{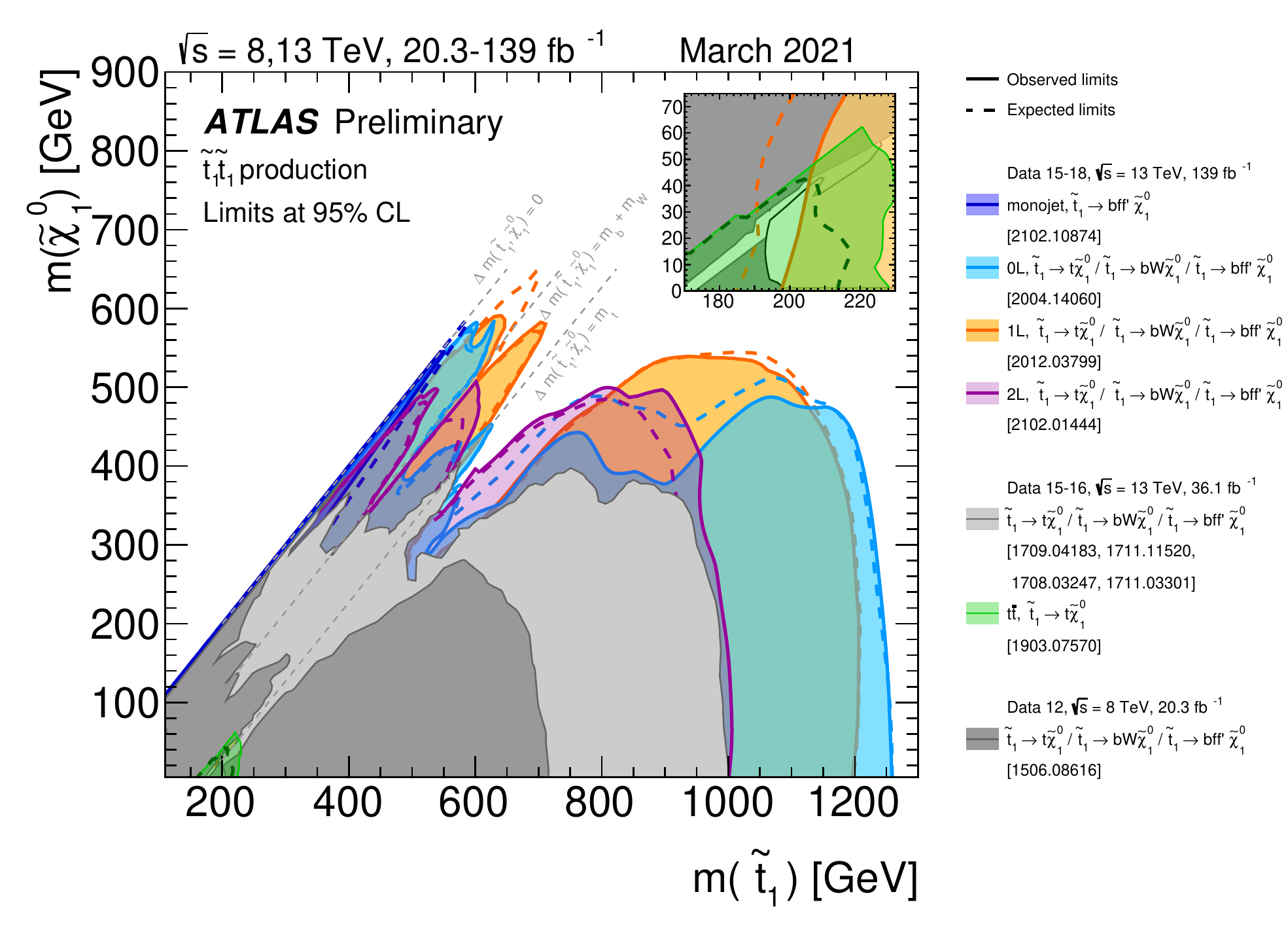}}
\subfigure{\includegraphics[width=0.4 \textwidth]{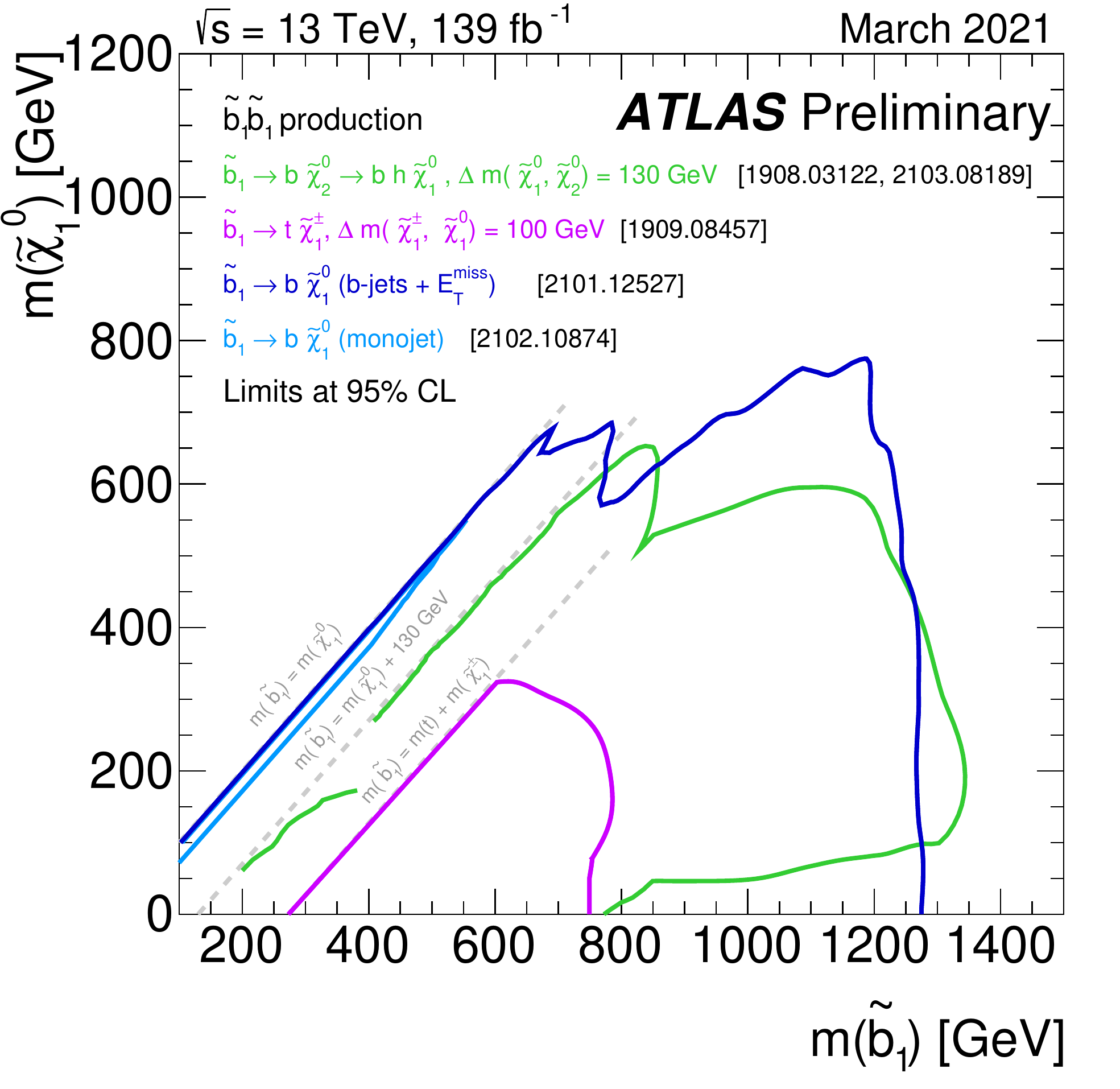}}
\caption{The limits on Stop masses (left panel) and sbottom masses 
(right panel) from the ATLAS collaboration at LHC.}
\label{stops and sbottoms}
\end{figure}

One of the most important searches conducted at LHC is for the third-generation squarks - the stops and sbottoms. The 
stop searches are in particular very interesting as they directly probe the radiative corrections to the Higgs mass
computation. Furthermore, information on their mixing angles can play an important role in understanding the
stable/unstable/metastable regions corresponding to charge and colour-breaking minima in the supersymmetric parameter
space. The limits on the stops are given in the left panel of Fig.\ref{stops and sbottoms} where limits from two-body
and three-body decay rates of the lightest stop are considered. The stops are ruled out between 200-1200 GeV. Similar results also exist for the sbottoms which are shown in the right panel of Fig.\ref{stops and sbottoms}. 

\begin{figure}[h!]
\centering    
\subfigure{\includegraphics[width=0.44\textwidth]{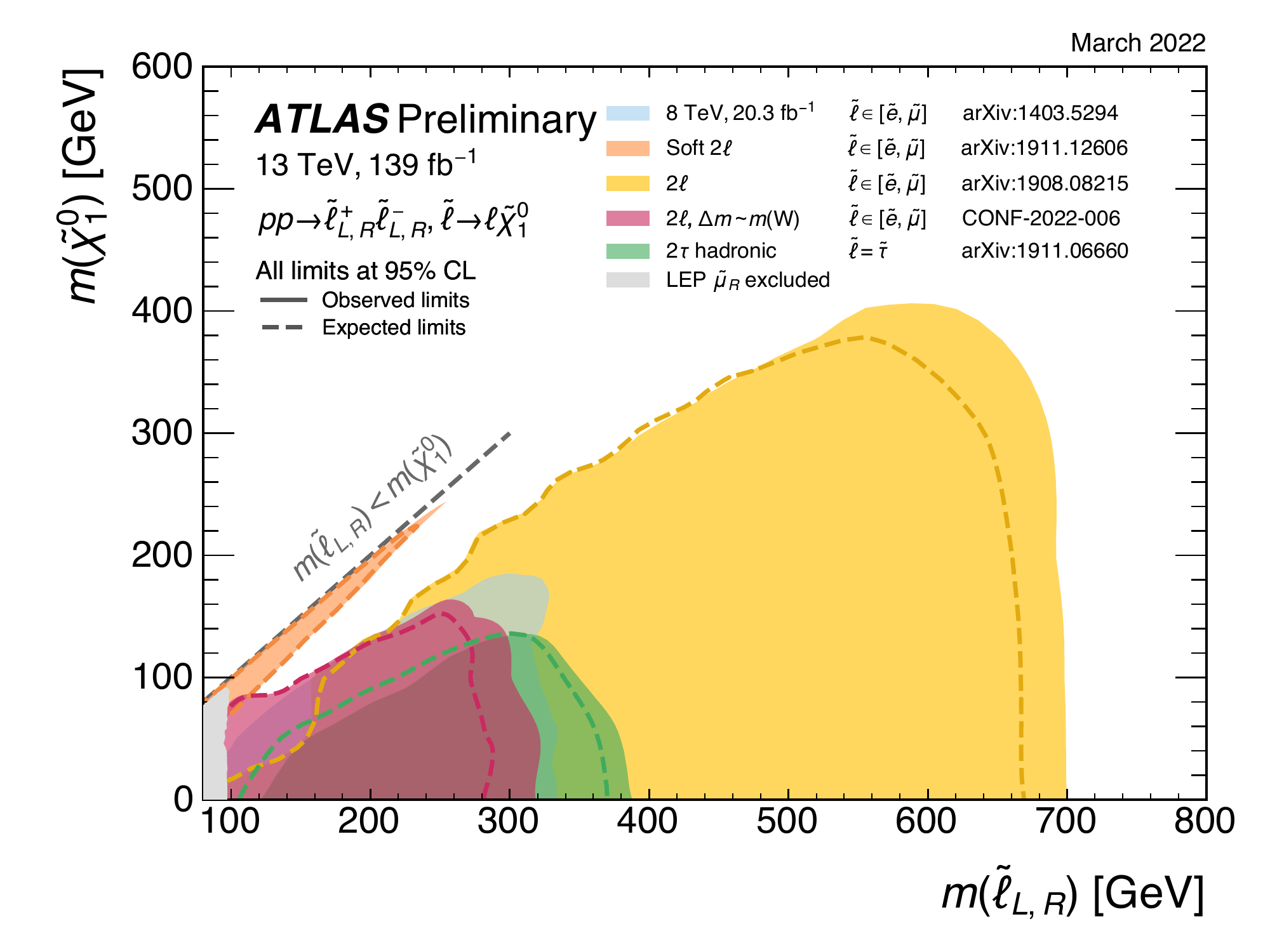}}
\subfigure{\includegraphics[width=0.51\textwidth]{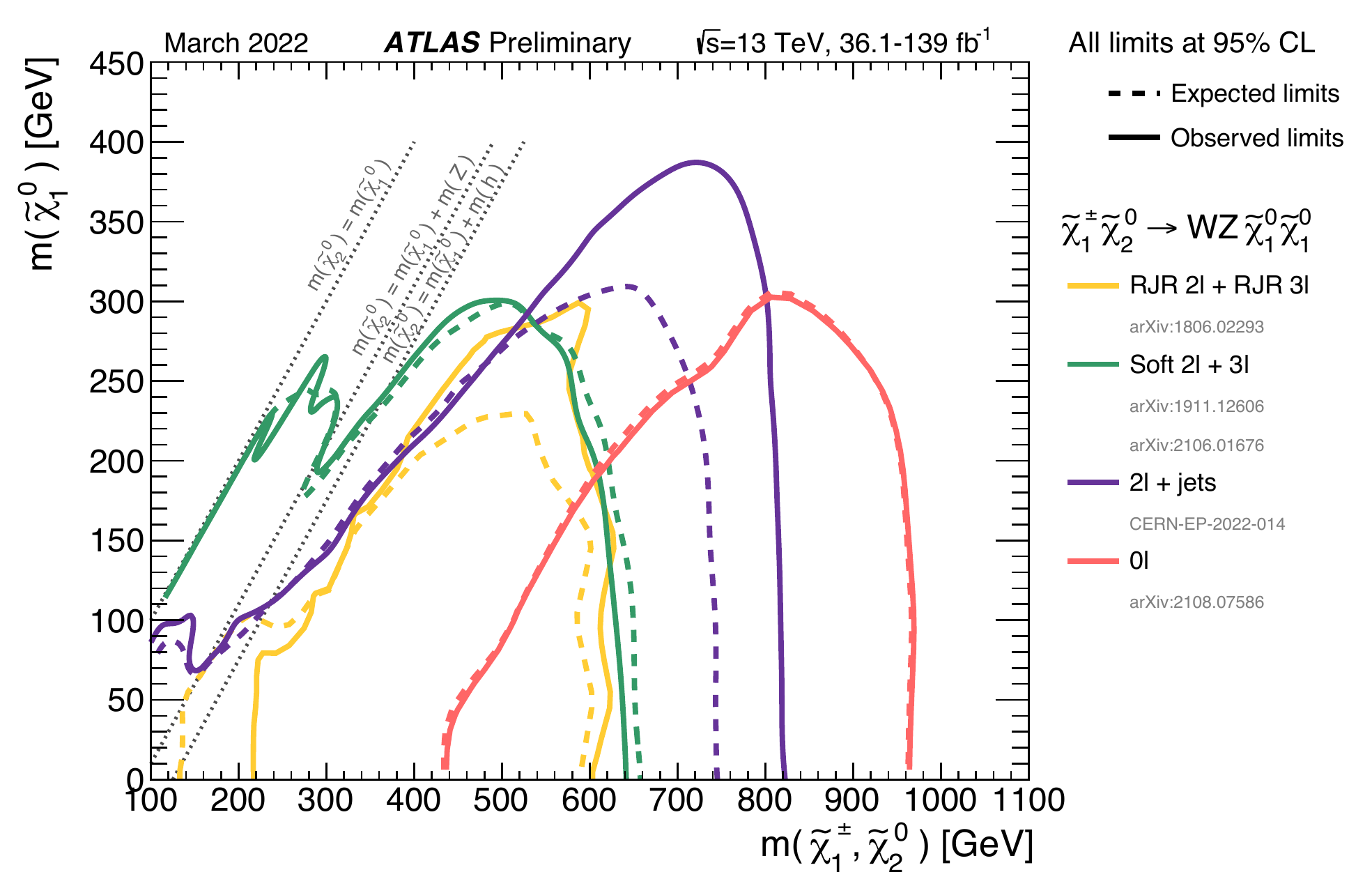}}
\caption{The limits from ATLAS collaboration on slepton masses (left panel) and wino-like neutralino mass (right panel).}
\label{sleptonwino}
\end{figure}

Finally, we look at the limits on weakly interacting particles which are dominantly produced by Drell-Yann like processes. Given that LHC is dominantly a hadron collider, the masses of slepton and other weakly interacting particles like charginos/neutralinos  probed are expected to
be weaker than that of gluinos or squarks. In figure \ref{sleptonwino}, limits are shown for sleptons (left panel) and for a pure wino-like neutralino (right panel).  It should be noted that most
of these limits are within simplified models of supersymmetry with branching fractions typically assumed to be at 100\% and so should be used with care. There could be large variations 
in other models. 

\subsubsection{Flavour Constraints}

 \begin{table}[h!]
\centering
\begin{tabular}{|c|c|c|}
\hline
Observable/Reaction & Measurement/Upper limits & Reference\\
\hline
	$\Delta M_{B^0}$ & $3.354\times 10^{-13}\mathtt{GeV}$ & \cite{Tanabashi:2018oca}\\
	$\Delta M_K$  & $3.484\times 10^{-15}\mathtt{GeV}$ & \cite{Tanabashi:2018oca} \\
	$\epsilon_K$ &  $1.596\times 10^{-3}$ & \cite{Tanabashi:2018oca} \\
	$a_{\mu}$ & $11659208.0\times10^{-10}$ & \cite{Bennett:2006fi} \\
	BR($\bar{B}\rightarrow X_s\,\gamma$) &  $3.29 \times10^{-4}$ & \cite{Lees:2012wg}\\
	Br($B\rightarrow X\,\mu\,\nu$) & 0.1086 &  \cite{Tanabashi:2018oca} \\
	Br($B\rightarrow X\, e\,\nu$)  & 0.1086&  \cite{Tanabashi:2018oca} \\
	Br($B_s\rightarrow \mu^+\,\mu-$) & $2.4\times 10^{-9}$&  \cite{Tanabashi:2018oca} \\
	Br($B^+\rightarrow \bar{D}^0\,l^+\,\nu$) & $2.27\times10^{-2}$ &  \cite{Tanabashi:2018oca} \\
	Br($B^-\rightarrow \bar{\pi}^0\,l^-\,\bar{\nu}$) & $7.80\times10^{-5}$ &  \cite{Tanabashi:2018oca} \\
	Br($K_L\rightarrow \mu^+\,\mu^-$) & $6.84\times 10^{-9}$ &  \cite{Tanabashi:2018oca} \\
	Br($K^+\rightarrow \mu^+\,\nu$) &  0.6356&  \cite{Tanabashi:2018oca} \\	
	\hline
	BR($\mu\rightarrow e\,\gamma$) & $ < 4.2\times10^{-13}$ & \cite{TheMEG:2016wtm}   \\
	BR($\mu\rightarrow$ e\, e\, e) & $< 1.0\times 10^{-12}$ & \cite{Bellgardt:1987du} \\
	BR($\tau\rightarrow \mu\,\gamma$)  & $<  4.4\times 10^{-8}$ & \cite{Amhis:2016xyh} \\
	BR($\tau\rightarrow e\,\gamma$)   & $< 3.3\times 10^{-8}$ & \cite{Amhis:2016xyh}  \\
	BR($\tau\rightarrow \mu\, \mu\, \mu$) & $< 2.1\times 10^{-8}$ & \cite{Amhis:2016xyh}   \\
	BR($\tau\rightarrow$ e\, e\, e) & $< 2.7\times 10^{-8}$ & \cite{Amhis:2016xyh}  \\
	BR($\tau\rightarrow \pi^0\,$ e) & $< 8.0 \times 10^{-8}$ & \cite{Miyazaki:2007jp}\\
	BR($\tau\rightarrow \pi^0\, \mu$) & $< 1.1 \times 10^{-7}$ & \cite{Aubert:2006cz}\\
	BR($\tau\rightarrow \rho^0\,$ e) & $< 1.8 \times 10^{-8}$ & \cite{Miyazaki:2011xe} \\
	BR($\tau\rightarrow \rho^0\, \mu$) & $ < 1.2 \times 10^{-8}$ & \cite{Miyazaki:2011xe}\\
	\hline
\end{tabular}
\caption{The experimental bounds on different flavour changing processes.}
\label{tab:flavorbound}
\end{table}

In the SM, flavor can change through either a neutral current or charged current. Flavour-changing neutral currents (FCNC) are absent in the SM at tree level by the GIM (Glashow-Illiopoulos-Maini) mechanism. In the charged current (CC) sector, flavour is violated in the SM by the well-known $V_{CKM}$ which is entirely fixed by experiments. A few flavour-changing transitions have been observed experimentally while many flavour-changing processes have upper bounds from different experiments. From table \ref{tab:flavorbound}, many upper bounds are several orders of magnitude above the SM expectations and these give indirect bounds on the scale of new physics.
		\begin{figure}[h!]
			\centering	\includegraphics[width=0.9\textwidth]{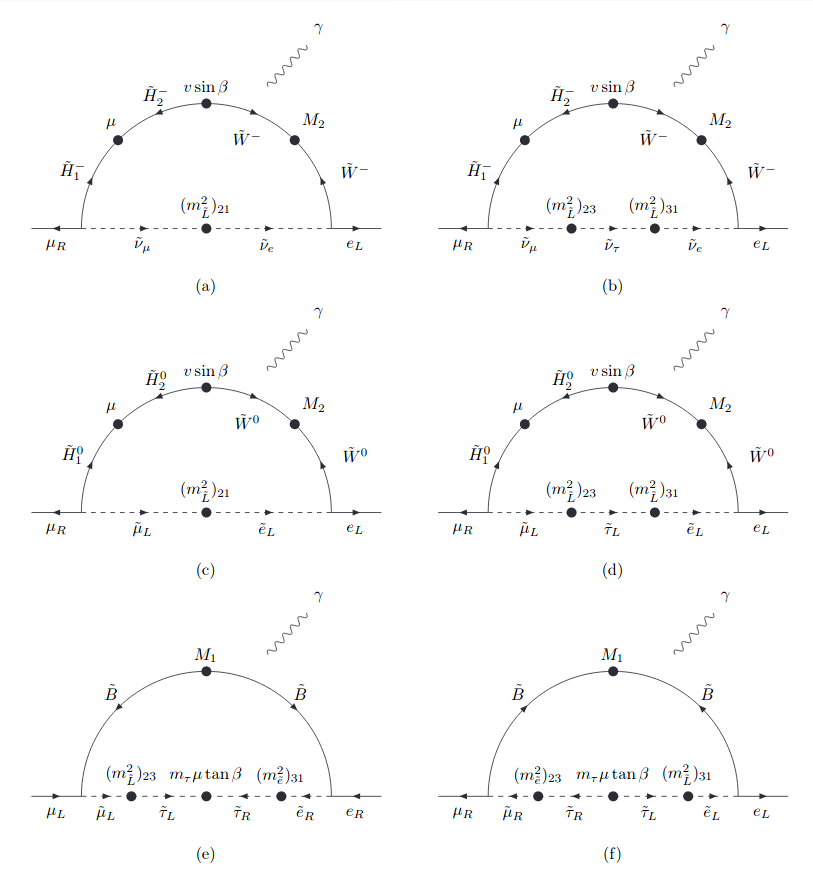}
			\caption{ One loop Feynman diagrams for $\mu \rightarrow e\, \gamma$ when the off-diagonal terms of slepton mass matrices are non-negligible, ref. \cite{Hisano:1998fj}}
			\label{fig:mueg}
		\end{figure} 
Let us assume a simple extension of the SM Lagrangian using an effective field theory approach for a new contribution to the flavour-changing process:
\begin{equation}
\mathcal{L}=\mathcal{L}_{SM} + \mathcal{L}_{eff}, \quad
\mathcal{L}_{eff} =\frac{1}{\Lambda^{d-4}}\sum_{i}^{}c_i(\mu)Q^i(\mu),
\end{equation}
where $Q^i$'s are a set of higher dimension operators which will contribute to flavour-changing processes; all these operators respect the SM gauge symmetry. Experiments put bounds on $c_i(\mu)/\lambda^{d-4}$, where $\Lambda$ is the scale of new physics. Both hadronic and leptonic FC processes tell us that new physics should be above 100 TeV for $\mathcal{O}(1)$ flavour violation.
In MSSM, FC source are soft-SUSY breaking terms for scalar masses:
\begin{eqnarray}
 \mathcal{L}_{soft} &= &m_{Q_{ii}}^2  \tilde{Q}_i^\dagger \tilde{Q}_i +
m_{u_{ii}}^2  \tilde{u^c}_i^\star \tilde{u^c}_i  + 
m_{d_{ii}}^2  \tilde{d^c}_i^\star \tilde{d^c}_i + 
m_{L_{ii}}^2  \tilde{L}_i^\dagger \tilde{L}_i + 
m_{e_{ii}}^2  \tilde{e^c}_i^\star \tilde{e^c}_i  \\ \nonumber 
&+& \left(\Delta_{i \neq j}^{u,d}\right)_{LL}  \tilde{Q}_i^\dagger \tilde{Q}_j + 
\left(\Delta_{i \neq j}^{u}\right)_{RR}  \tilde{u^c}_i^\star \tilde{u}^c_j  +
\left(\Delta_{i \neq j}^{d}\right)_{RR}  \tilde{d^c}_i^\star \tilde{d}^c_j  \\ \nonumber
&+& \left(\Delta_{i \neq j}^{l}\right)_{LL}  \tilde{L}_i^\dagger \tilde{L}_j + 
 \left(\Delta_{i \neq j}^{l}\right)_{RR}  \tilde{e}_i^\star \tilde{e^c}_j  + \ldots .
\end{eqnarray}
and trilinear scalar couplings:
\begin{equation}
\mathcal{L}_{soft} = A^u_{ij}\tilde{Q}_i\tilde{u}^c_jH_2 + A^d_{ij}\tilde{Q}_i\tilde{d}^c_jH_1 + A^e_{ij}\tilde{L}_i\tilde{e}^c_jH_1.
\end{equation}
The soft mass term $\Delta_{ij}(i\neq j,LL/RR)$ and the trilinear scalar terms $\Delta_{ij}\propto A_{ij}(\Delta_{ij}\propto A_{ij} \mbox{for}, i\neq j,LR)$ can violate flavour in both squark and slepton sector. Furthermore, all these couplings can be complex and will give CP violation in addition to the CKM phase. These terms can contribute dominantly compared to the SM amplitudes to various flavour-violating processes at the weak scale, such as flavour oscillations ($K^0\leftrightarrow \bar{K}^0$), flavour-violating decays like $b\rightarrow s \gamma$ and even in flavour violating decays which do not have any SM contributions like $\mu \rightarrow e \gamma$, see fig. (\ref{fig:mueg}).
 
To analyze the phenomenological impact of flavour-changing processes on flavour-violating terms, a useful tool called Mass Insertion approximation can be used. In this approximation, the flavour changing is encoded in non-diagonal sfermion propagators with flavour diagonal gaugino vertices. These propagators are then expanded, assuming that the flavour-changing parts are much smaller than the flavour diagonal ones. Then we define flavour violating parameters, $\delta_{ij}^f\equiv \Delta_{ij}^f/m_{\tilde{f}}^2$, known as mass insertions; where $\Delta_{ij}^f$ are the flavour-violating off-diagonal entries appearing in the f=(u,d,l) sfermion mass matrices and $m_{\tilde{f}}^2$ is the average sfermion mass. The mass insertions are further subdivided into LL/RR/LR/RL types, labelled by the corresponding SM fermions' chirality.

\begin{table}[h!]
    \centering
    \begin{tabular}{||c|c|c|c||}
    \hline
        Type of $\delta^l$ & $\mu \rightarrow e\,\gamma$ ($\delta_{12}$) & $\tau \rightarrow \mu\,\gamma$ ($\delta_{23}$) & $\tau \rightarrow e\,\gamma$ ($\delta_{13}$) \\
        \hline
         LL & $6\times 10^{-4}$ & 0.12 & 0.15\\
         LR/RL & $1 \times 10^{-5}$ & 0.03 & 0.04\\
         \hline
    \end{tabular}
    \caption{The bounds are obtained by scanning over $m_0 < 380$ GeV, $M_{1/2} < 160$ GeV, $|A_0|\leq 3m_0$ and $5<\tan{\beta}<15$ \cite{Ciuchini:2007ha}.}
    \label{tab:leptondelta}
\end{table}

\begin{table}[h!]
    \centering
    \begin{tabular}{||c|c|c|c|c||}
    \hline
      $\delta_{ij}^d$   & LL & LR & RL & RR \\
      \hline
      12     &  $1.4\times 10^{-2}$ & $9.0\times 10^{-5}$ & $9.0\times 10^{-5}$ & $9.0\times 10^{-3}$\\
      13 & $ 9.0\times 10^{-2}$ & $1.7\times 10^{-2}$ & $1.7\times 10^{-2}$ & $7.0\times 10^{-2}$\\
      23 & $0.16$ & $4.5\times 10^{-3}$ & $6.0\times 10^{-3}$ & 0.22\\
      \hline
    \end{tabular}
    \caption{The bounds are obtained by scanning over $m_0 < 380$ GeV, $M_{1/2} < 160$ GeV, $|A_0|\leq 3m_0$ and $5<\tan{\beta}<15$ \cite{Ciuchini:2007ha}. The constraints on $(\delta^d_{12})_{AB}$ (AB=LL,RR,LR,RL) are taken from the measurements of $\Delta M_K, \epsilon$ and $\epsilon'/\epsilon$. The $(\delta^d_{13})_{AB}$ constraints are taken from the measurements of $\Delta M_{B_d}$ and $2\beta$ and the $(\delta^d_{23})_{AB}$ constraints are taken from the measurements of $\Delta M_{B_s}$, $b \rightarrow s\, \gamma$ and $b \rightarrow s\, l^+\,L^-$. }
    \label{tab:quarkdelta}
\end{table}
Tables \ref{tab:quarkdelta} and \ref{tab:leptondelta} are lists of constraints on $\delta$'s from different hadronic and leptonic flavour-changing processes. In both sectors, $\delta$'s are suppressed to $10^{-5}$ for SUSY scale near the EW scale. Let's take an example to understand these bounds. Assume that flavor violating term is only $(\delta^l_{12})_{LL}$ (all others are zero), and all SUSY-particles have same mass, then branching fraction of $\mu \rightarrow e\, \gamma$ is \cite{Hisano:2001qz}:
\begin{equation}
Br(\mu \rightarrow e \gamma) \simeq 1.18 \times 10^{-6}\left(\frac{\tan{\beta}}{15}\right)^2 \left(\frac{1\,\mbox{TeV} }{m_{SUSY}}\right)^4(\delta^l_{12})_{LL}^2.
\end{equation}
This requires either for SUSY particle to have masses around 50-60 TeV with flavour violation $\mathcal{O}(1)$ or a 1 TeV scale SUSY spectrum with suppressed flavour violating terms $(\delta^l_{12})_{LL}<6\times 10^{-4}$. LR bounds further push SUSY masses to be greater than 100 TeV for $\mathcal{O}(1)$ flavour violation. The flavour constraints are 50-60 times stronger than LHC bounds with flavour violation in supersymmetry. These limits show that the flavour-violating terms should be typically at least a couple of orders of magnitude suppressed compared to the flavour-conserving soft terms. The flavour problem requires heavy-scale SUSY of around 100 TeV, which decouples soft masses from weak-scale physics, or some alignment mechanisms. As we can see from tables \ref{tab:leptondelta} and \ref{tab:quarkdelta}, flavour violation in the first two generations of sfermion masses is more constrained from experiments than the third generation. Then the decoupling of the first two generations can avoid bounds from various experiments. The other way to avoid these bounds is to suppress flavour-violating entries to zero through some symmetries or by some flavor-suppressing SUSY breaking mechanism \cite{Dudas:2019gxd}.

\subsubsection{Minimal Flavor Violation}

An important point is that if we set all the flavour-violating off-diagonal entries to zero
through some mechanism or by choosing an appropriate supersymmetry breaking 
mechanism,  the contribution from the supersymmetric 
sector to flavour violation will not be completely zero. This is because CKM (Cabibbo-Kobayashi-Masakawa)
 matrix will induce non-trivial flavour-violating interactions between the SM 
 fermion and its supersymmetric partner.  One of the strongest constraints, in this
case, comes from BR$(b\to s + \gamma)$ which has been measured very precisely
by the experimental collaborations (with an error of about 5\% at the one sigma level). 
The present numbers are as follows \cite{Misiak:1997ei}
\begin{eqnarray}
BR(b \to s+ \gamma)^{exp} &=& (3.43 \pm 0.21 \pm 0.07) \times 10^{-4} \nonumber \\ 
BR(b \to s+ \gamma)^{SM} &=& (3.36 \pm 0.23) \times 10^{-4} .
\end{eqnarray}
Given the closeness of the Standard Model expectation to the experimental number, any
new physics should either be very heavy  such that it's contributions to this rare process
are suppressed or should contain cancellations within its contributions such that the total
SM+ New physics contribution is close to the experimental value. Both these scenarios
are possible within the MSSM. If supersymmetric partners are heavy $\geq$ a few TeV, 
then their contributions to $b\to s+\gamma$ are highly suppressed. On the other hand, 
it is possible that the dominant  contributions from  charged Higgs  and the 
chargino diagrams cancel with each other (they come with opposite sign) for a large region
of the parameter space.  The general class of new physics models which do not introduce
any new flavour violation other than the one originating from the CKM matrix in the Standard
Model come under the umbrella of  ``Minimal Flavour Violation" \cite{DAmbrosio:2002vsn}. 

\subsubsection{Anomalous magnetic Moment of Muon $(g-2)_{\mu}$} 

The anomalous magnetic dipole moment of the muon, commonly known as the $g-2$ anomaly, can be considered as a hint for physics Beyond the Standard Model (BSM), particularly supersymmetry. The E821 experiment at BNL \cite{Muong-2:2006rrc} first showed a discrepancy compared with the theoretical predictions, at a significance of $\sim 3.7 \sigma$. This was confirmed in 2021 by the E989 experiment at Fermilab \cite{Muong-2:2021ojo,Muong-2:2021ovs,Muong-2:2021vma}, which yielded a measurement of $a_{\mu}^{\rm{EXP}} = 116592061(41)\times10^{-11}$. The Fermilab measurement, in comparison with the latest theoretical SM prediction of $a_{\mu}^{\rm{SM}} = 116591810(43)\times10^{-11}$ \cite{Aoyama:2020ynm}, further increases the discrepancy to $4.2\sigma$
\begin{equation}\label{eq:DeltaaExp}
\Delta a_{\mu} = a_{\mu}^{\rm{EXP}} - a_{\mu}^{\rm{SM}} = 251(59) \times 10^{-11}.
\end{equation}

In recent times, some lattice collaborations \cite{Borsanyi:2020mff,Ce:2022kxy} appear to agree with the experimental measurements, thus eliminating the anomaly. Therefore, it seems that further confirmation is needed. The upcoming runs of the E989 experiment are expected to lower the experimental uncertainty by a factor of 4, whereas the E34 collaboration at J-PARC \cite{Saito:2012zz,Mibe:2011zz,Abe:2019thb} is planning on measuring $a_{\mu}$ using a completely different method. As for the SM predictions, various theory and lattice groups are expected to present updated results from those in \cite{Borsanyi:2020mff,Ce:2022kxy}.

Muon anomalous magnetic moment has long been considered as a signature of 
supersymmetry\cite{Chattopadhyay:1995ae,Moroi:1995yh}. However, taken together with the recent constraints (discussed in Section 1), from the Higgs mass, LHC results and so on, the present supersymmetric explanations are significantly different compared to the older ones. See for example, 
the discussion in \cite{Baum:2021qzx,Endo:2021zal}. The key in both explanations is to generate a small $\mu$ parameter.


\subsubsection{Dark Matter} 

In 1933, Zwicky \cite{Zwicky:1937zza,Zwicky:1933gu} analyzed the motion of galaxies in the Coma cluster. Clusters are bound systems of galaxies, and their masses can be calculated using the Virial theorem. He showed that the mass necessary to explain the unity of the Coma cluster was much higher than the luminous mass of the cluster. It was the first evidence of non-luminous matter. Later, V. Rubin measured the rotation curves of galaxies in 1969 \cite{Rubin:1970zza}, and the flatness of rotation curves at a larger distance could not be explained only by luminous matter. They needed a large amount of non-luminous matter to explain it. Later this non-luminous matter was named “Dark matter” and since then, various experimental observations have confirmed the existence of dark matter, like the kinematics of virial systems and rotating spiral galaxies, gravitational lensing of background objects, gravitational lensing and X-ray observations from the Bullet Cluster, and studies of CMB. The Planck collaboration has precisely measured the dark matter relic density, $\Omega h^2=0.120\pm0.001$ at $68\%$ C.L. \cite{Aghanim:2018eyx}, using CMB anisotropies. The most popular candidate for dark matter is Weakly Interacting Massive Particle (WIMP), which satisfies the relic abundance of DM.  There are majorly three ways to detect dark matter
\begin{itemize}
\item Indirect detection: Indirect searches from DM annihilation and decay are conducted using various satellites and telescopes (Fermi-LAT, Planck, PAMELA, AMS, H.E.S.S.). We have to deal with a lot of astrophysical backgrounds in these experiments.
\item Direct detection: In these experiments, we try to detect the scattering of dark matter from the halo of target nuclei on Earth to calculate mass and coupling of dark matter with SM. Heavily shielded underground detectors are used to detect dark matter. These detectors measure the amount of energy deposited by DM during scattering with target nuclei (DAMA, LUX, CDMS, CoGeNT, CRESST, PandaX, XENON1T, etc.).
\item Searches in colliders: Limits on dark matter are obtained from searches of dileptons, dijet resonances, monojet, single photons with missing transverse energy, at LHC and LEP.
\end{itemize}
\begin{figure}[h!] 
\centering    
\subfigure{\includegraphics[width=0.47\textwidth]{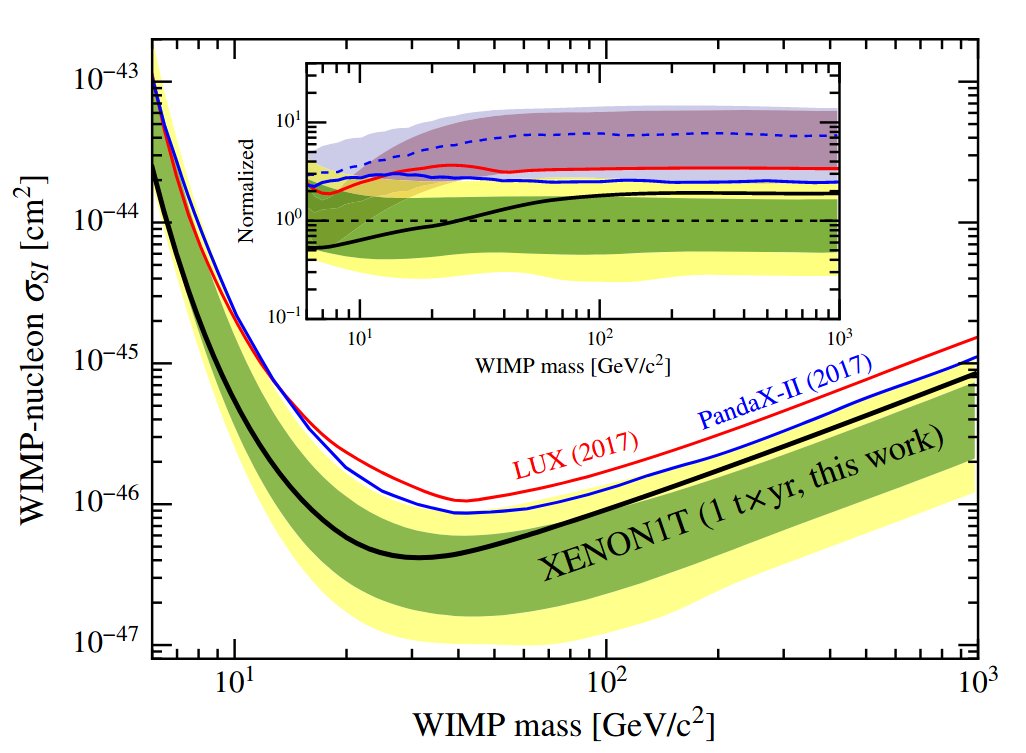}}
\subfigure{\includegraphics[width=0.485\textwidth]{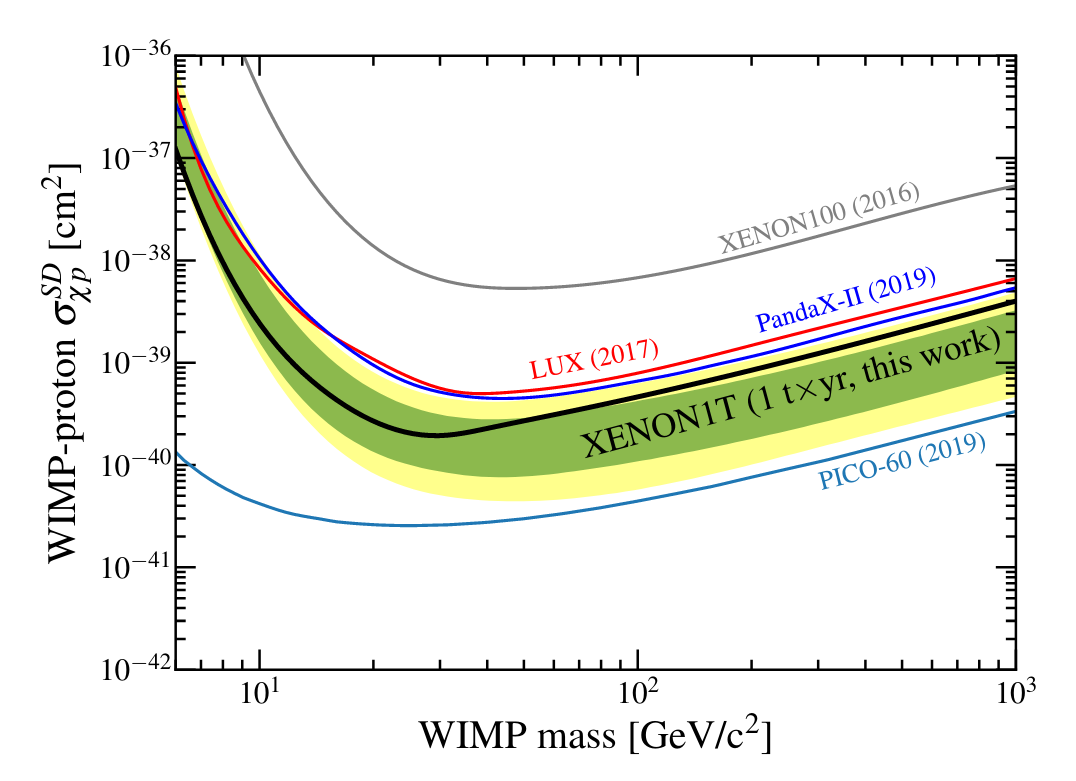}}
\caption{\textbf{left:} Spin-independent WIMP nucleon cross-section vs WIMP mass \cite{Aprile:2018dbl}. Xenon1T gives the strongest bound. \textbf{right:} Spin-dependent WIMP nucleon cross-section vs WIMP mass \cite{Aprile:2019dbj}.PICO-60 $C_3F_8$ \cite{Amole:2019fdf} gives the strongest bound.
}
\label{fig:SISDWIMP}
\end{figure}
The exciting phenomenological feature of R-parity conserving SUSY is that the LSP is stable and is a good candidate for DM. The nature of LSP depends on the SUSY breaking mechanism. In some SUSY models like SUGRA, the LSP is the neutralino (pure gaugino, pure higgsino, or mixed state of gaugino and higgsino) whereas in other SUSY models like GMSB, the LSP is gravitino. Various experimental bounds like the LHC searches for missing energies, cosmological data, and direct detection of dark matter are hunting for SUSY dark matter from all directions. For example, gravitino dark matter has strong bounds from cosmological data, whereas, neutralino dark matter has significant constraints from direct detection and LHC searches. The dark matter density is also a big constraint on LSP masses. Consider the lightest neutralino which has the following form
\begin{equation}
\chi^0_1 = N_{\tilde{B}_1} \tilde{B}^0 + N_{\tilde{W}_1 } \tilde{W}^0 + N_{(\tilde{H}_u)_1} \tilde{H}_u^0 +
N_{(\tilde{H}_d)_1} \tilde{H}_d^0,
\end{equation}
where N's are mixing parameters, and look at the various possible compositions of the LSP to satisfy the relic density.\\
\noindent
(a) \textit{Pure bino:} The annihilation cross-section is given by \cite{ArkaniHamed:2006mb}
\begin{equation}
\langle \sigma_{\chi} v \rangle = \frac{3 g^4 \tan\theta_W^4 r(1+r^2)}{2 \pi m_{\tilde{e}_R}^2 x (1 +r)^4},
\end{equation}
where $x= M_1/T$ the mass of the bino over the temperature and  $r \equiv M_1^2/m_{\tilde{e}_R}^2$, $\theta_W$ is the weak mixing angle, or the Weinberg angle. The relic density in this case is given by 
\begin{equation}
\Omega_{\tilde{B}}h^2 = 1.3 \times 10^{-2}  \left( \frac{  m_{\tilde{e}_R}}{100~ \rm{GeV}}\right)^2 \frac{ (1+r)^4}{r^2 (1+r)^2} \left(1+ 0.07 \ln\frac{\sqrt{r} ~~100 ~\rm{GeV}}{m_{\tilde{e}_R}} \right).
\end{equation}
The above relic density is typically large for a reasonable range of parameters. One thus typically invokes co-annihilating partners which have a mass that is very close to that of the bino, thereby increasing the cross-section and bringing down the relic density to observed values. \\
\noindent
(b) \textit{Pure wino:} The annihilation cross-section of the dark matter particle is proportional to $g^4$, the weak coupling:
\begin{equation}
\langle \sigma_{\chi} v \rangle  =  \left( \frac{3 g^4}{16 \pi M_2^2} \right),
\end{equation}
where $M_2$ stands for the wino mass. The relic density is approximately given by
\begin{equation}
\Omega_{\tilde{W}}h^2 \sim 0.13   \left( \frac{M_2}{2.5~ \rm{TeV}}\right)^2.
\end{equation}
The observed relic density requires a heavy neutralino of the order of 2.3 TeV. \\
\noindent
(c) \textit{Pure higgsino:} The annihilation cross-section of the dark matter particle  is given by
 \begin{equation}
\langle \sigma_{\chi} v \rangle = \frac{3 g^4}{512 \pi \mu^2} \left( 21 + 3 \tan\theta_W^2  + 11 \tan\theta_W^4 \right).
\end{equation}
The relic density, in this case, is given by 
\begin{equation}
\Omega_{\tilde{H}}h^2 \sim 0.10   \left( \frac{\mu}{1~ \rm{TeV}}\right)^2.
\end{equation}
A neutralino of 1 TeV is required to satisfy the relic density.\\ In summary, a pure bino neutralino can be light but it requires co-annihilating partners (or some other mechanism) to give the correct relic density, whereas both a pure higgsino and a pure wino would have to be close to a TeV or larger. Admixtures of various components (known as well-tempered dark matter\cite{ArkaniHamed:2006mb}) can, however, give the right relic density. 
		\begin{figure}[h!]
			\centering
			\includegraphics[width=0.7\textwidth]{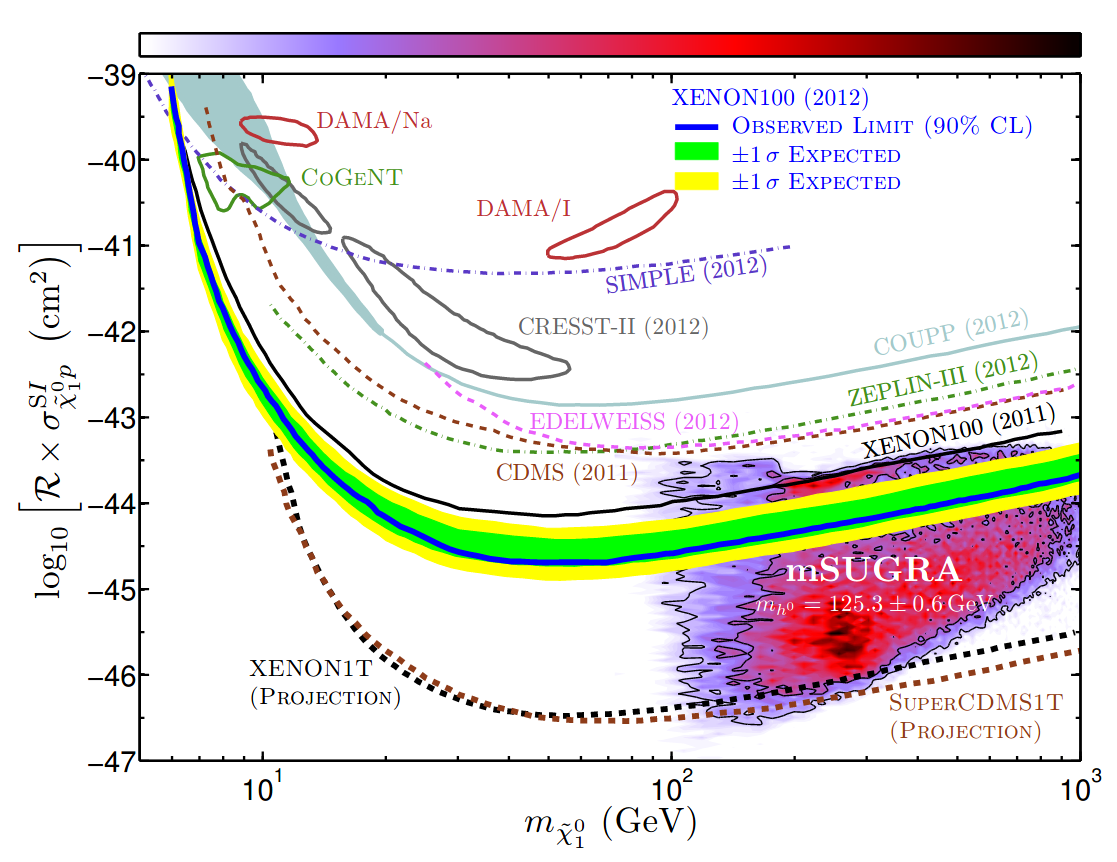}
			\caption{The spin-independent neutralino-proton cross-section as a function of the neutralino mass for the model mSUGRA. It is seen that most of the parameter space is ruled out by current XENON-1T data \cite{Nath:2012nh}.}
			\label{fig:sugradrak}
		\end{figure}
There has been tremendous progress in the direct detection experiments of  WIMP dark matter. The spin-independent WIMP-nucleon cross-section is constrained to be smaller than a few $\times ~10^{-47}~\rm{cm}^{-2}$ for DM masses of $\mathcal{O}(20-100)$ GeV \cite{Aprile:2018dbl,Aprile:2019dbj,Amole:2019fdf}, shown here for DM masses up to 1 TeV (fig. \ref{fig:SISDWIMP}). Many supersymmetric models with TeV scale supersymmetry predict a WIMP of this mass and cross-sections of this order. Taken together with relic density requirements, they strongly constrain most regions of supersymmetric parameter space. It leaves only some special regions like coannihilation regions, funnel regions (for example, see \cite{Bagnaschi:2017tru,Costa:2017gup}), and those corresponding to well-tempered dark matter. In fig. (\ref{fig:sugradrak}), we see direct detection measurements rule out almost all parameter space of DM in mSUGRA models (usually LSP is bino in this model). Similarly, the cosmological bound on gravitino mass, together with Higgs mass and direct searches at LHC, excludes minimal gauge mediation with high reheating temperatures. LHC searches also have not found any signal of dark matter and constraints neutralino mass to $>600$ GeV. \\
Therefore, dark matter studies of supersymmetric theories need either heavy supersymmetry or a SUSY breaking mechanism that gives a well-tempered dark matter or more coannihilation and funnel region.



\section{Supersymmetric Spectra and Models}

\begin{figure}[h!] 
\centering    
\includegraphics[width=\textwidth]{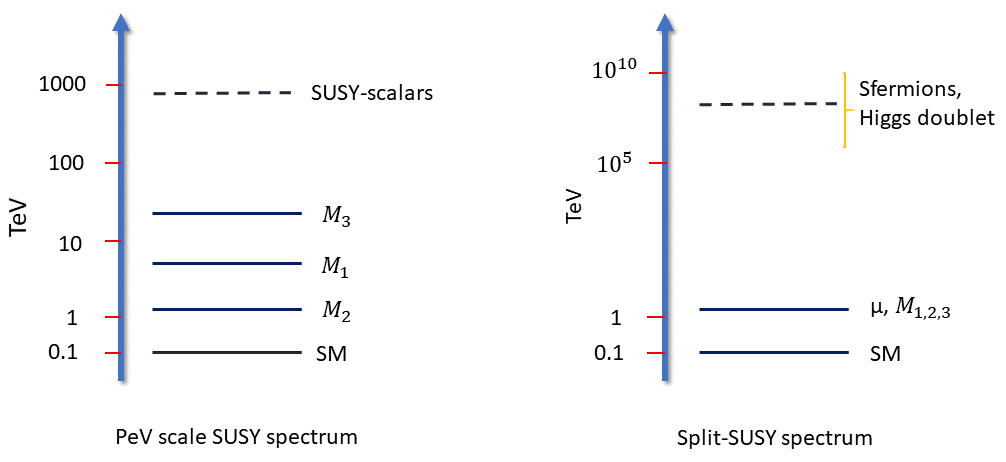}
\caption{Examples of heavy scale SUSY spectrum. \textbf{left:} Spectrum of PeV scale SUSY proposed by \cite{Wells:1997af}. \textbf{right:} Spectrum of Split-SUSY proposed by \cite{ArkaniHamed:2004fb,Giudice:2004tc}}
\label{fig:heavysusy}
\end{figure}

Most of the supersymmetric analysis earlier has been mainly focussed on  models which can be broadly classified as (a) gravity-mediated models like mSUGRA/cMSSM, Anomaly mediated supersymmetry breaking, moduli-mediation, etc.. and (b) gauge-mediated supersymmetry breaking and its variants. Both scenarios typically have strong correlations among supersymmetric spectrum masses and couplings at the weak scale. This is typical because very few parameters determine the entire mass spectrum and couplings. After the Higgs discovery and LHC results, the kind of models discussed above are under severe stress and perhaps might not be testable in present colliders like LHC unless for special regions of the parameter space. 

While most of the present analyses do happen in the phenomenological 
MSSM currently, one can still think of possible spectrum/models which 
are most likely suggested by the data. A first approach is to consider 
the degenerate spectrum which will kinematically suppress the final states
of susy particle cascade decays to be below the threshold energies of the collider.  The other classes of models are hierarchical supersymmetric
spectrums.

As we discussed above,  data from various experiments (see table \ref{tab:bounds}) give us hints of hierarchical supersymmetry. In heavy-scale supersymmetry, the most unpleasant aspects of low-scale supersymmetry like flavour and CP violation constraints, fast proton decay constraints, Higgs mass constraints, and LHC constraints are eliminated. It does not affect SUSY's nice features like gauge couplings unification.
\begin{table}[h!]
    \centering
    \begin{tabular}{||c|c||}
    \hline
      Constraints   & Bounds on SUSY Spectrum  \\
      \hline
      Flavor and CP violation   & $1^{st}$ two generations $>100$ TeV\\
      Proton decay in minimal SUSY-SU(5) & sfermions $>50$ TeV\\
      LHC & $> 2.2$ TeV\\
      Higgs & $> 2$ TeV or $> 8$ TeV with $A=0$\\
      \hline
    \end{tabular}
    \caption{Summary of data coming from various direct and indirect experiments.}
    \label{tab:bounds}
\end{table}

It is possible for SUSY to be heavy, all the way upto GUT scale, and to be completely decoupled from low-scale physics. In this case, it does not have a phenomenological interest. Other possibilities are, see fig. (\ref{fig:heavysusy}), scalars of supersymmetry are at heavy scale, but gauginos are at the TeV scale (chiral symmetries can protect them). They will be seen in future experiments (33 TeV, 100 TeV, and high luminosity collider).
\begin{figure}[h!] 
\centering    
\includegraphics[width=\textwidth]{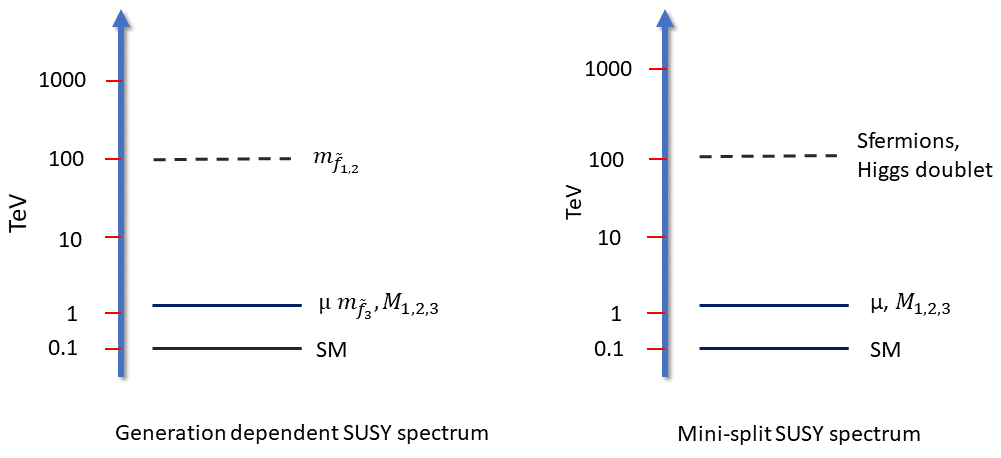}
\caption{Spectra of SUSY models which have scalar masses near 100 TeV.}
\label{fig:miniheavy}
\end{figure}

The Higgs mass is correlated to scalar masses, and for Higgs mass 125.01 GeV, the scalar masses should be in the range of 10 TeV to $10^5$ TeV \cite{Arvanitaki:2012ps}. For $\tan{\beta}\sim 3$, scalar masses should be less than 100 TeV or at least the third generation of scalar masses should be light (few TeV). There are models like mini-split \cite{Arvanitaki:2012ps} in which all scalars are around 100 TeV, and gauginos are around the TeV scale, extensively studied in the literature. Similarly, there are models where generation-dependent SUSY spectrum is considered \cite{Randall:2012dm,Cohen:1996vb,Dimopoulos:1995mi}, in which 1$^{st}$ two-generations are heavy but third generation and gauginos are around a few TeV. These models have very nice phenomenology and are accessible at future experiments. 

\section{Outlook}

Supersymmetry is at crossroads currently. While the experimental support for supersymmetry is currently very scant, or even absent, theoretically perhaps, supersymmetric standard models are one of the well-motivated models of physics beyond standard models. One can even push further and claim that the discovery of the Higgs and the anomaly of the muon $g-2$ are indirect indications of supersymmetry. Only future experiments can tell, whether faith in supersymmetry is the right direction to go. 
Till then, we can keep our fingers crossed and hope for the best.

\section{Acknowledgements}
S.K.V. thanks SERB Grant CRG/2021/007170 ``Tiny Effects from Heavy New Physics" from the Department of Science and Technology, Government of India.

\bibliographystyle{unsrt}
\bibliography{susyreview.bib}

\end{document}